# In situ nonlinear Rayleigh wave technique to characterize the tensile plastic deformation of stainless steel 316L

Changgong Kim, Kathryn H. Matlack

**Abstract**

The acoustic nonlinearity parameter $\beta$ is sensitive to dislocation parameters, which continuously change during plastic deformation. Dislocation-based damage in structures/components is the source of the failure; thus, $\beta$ has been studied as a metric for non-destructive evaluation. This work consists of two parts: the development of an in situ experimental setup for nonlinear Rayleigh wave measurements, and characterization of the dependence of $\beta$ on applied stress at different levels of initial plastic strain. First, we introduce an experimental setup and methods for repeatable in situ nonlinear ultrasonic measurements. Details on design considerations and measurement schemes are provided. In the second part, $\beta$ was measured *in situ* during an incremental monotonic tensile test. The measured $\beta$ monotonically decreases with plastic strain, but it is relatively insensitive to the applied stress during elastic deformation. This result highlights three aspects of the evolution of $\beta$, which have not been sufficiently emphasized in prior work: the apparent insensitivity of $\beta$ to the applied stress during elastic deformation, decreasing $\beta$ with plastic deformation, and the saturation of $\beta$. We attribute the trend of decreasing $\beta$ to a scaling of $\beta$ with monopole loop length during plastic deformation, which depends on initial microstructure. The saturation of $\beta$ at 1.8% coincides with a planar-to-wavy transition of dislocation structures. The in situ nonlinear ultrasonic experimental method presented in this work is significant as the in situ results can provide broader insights on $\beta$ and dislocation-based damage evolution than ex situ measurements alone.

## 1. Introduction

When materials plastically deform, dislocations are generated to accommodate the imposed damage. As the damage accumulates, the number density of dislocations increases, and they evolve into more complex but organized configurations. Nonlinear ultrasound (NLU) is sensitive to the changes in dislocation structures, which enables the nondestructive evaluation of dislocation-based damage by measuring the nonlinear ultrasonic response of dislocations. The acoustic nonlinearity parameter $\beta$ is often used to quantify the nonlinearity of dislocation motions. The nonlinearity of dislocations depends on dislocation parameters, which continuously change during damage accumulation regardless of the damage mechanisms. Thus, the sensitivity of the NLU technique to dislocation parameters makes it a potentially useful NDE tool for monitoring structures/components exposed to dislocation-based damage.

According to the theoretical models, $\beta$ depends on material constants, internal stress, and dislocation parameters, e.g., dislocation density, monopole loop length, and dipole height[1–4]. The dependence of $\beta$ on dislocation parameters has been extensively studied for damage mechanisms such as monotonic tension[5,6], fatigue[2,5,7], and precipitation[8,9]. The results reported in the prior work confirm that the evolution of $\beta$ correlates to the changes in dislocation parameters. However, there has been little work on the effect of internal stress on $\beta$. While Shui et al. measured $\beta$ as a function of the applied stress during monotonic tension tests[10], the effect of the applied stress was not discussed in detail.

Cash et al. were the first who highlighted the significance of the internal stress[4]. They updated Cantrell's models[2] to include the stress-dependence of $\beta$, and their models highlight two aspects of the evolution of $\beta$ during plastic deformation: the possibility of negative $\beta$ and the dependence of $\beta$ on the internal stress and dislocation configurations. Recently, Gao et al. also developed a dislocation pile-up model that predicts negative $\beta$[11]. The negative contribution to $\beta$ is especially noteworthy since it can cancel out positive contributions and underestimates $\beta$. The significance of negative $\beta$ is that the measured $\beta$ can decrease with damage accumulation, which contrasts with typical findings in prior work: a monotonic increase in $\beta$ with damage accumulation. However, the decrease in $\beta$ during plastic deformation has been rarely reported.

The stress dependence of $\beta$ can be experimentally studied using in situ NLU measurements. The in situ NLU measurements allow to change the stress term while keeping other model parameters constant. There is a few work on in situ NLU measurements. For example, Shui et al.[10] made in situ $\beta$ measurements using longitudinal waves on magnesium-aluminum alloy and Kim et al.[12] used Rayleigh waves to measure the dependence of $\beta$ on creep and cyclic loading of concrete. However, prior work does not highlight the subject of this work: in situ setup, measurement procedures, and, above all, the dependence of $\beta$ on the applied stress.

The first goal of this work is to develop *in situ* Rayleigh wave measurement fixtures and procedures for repeatable $\beta$ measurements while a specimen is loaded. Compared to longitudinal wave measurements, nonlinear Rayleigh wave measurements can be more sensitive to the setup because the propagating wave transmits through multiple interfaces, e.g., transducer to wedge and wedge to the material. Further, *in situ* measurements during mechanical testing requires special considerations since measurements are done while a specimen is vertically mounted in the load frame. In the first part of work, we introduce *in situ* setup and measurement procedures for repeatable *in situ* NLU measurements.

The second goal of this work is to characterize the dependence of $\beta$ on applied stress in the elastic and plastic regimes, using our *in situ* nonlinear Rayleigh wave measurement setup. The results highlight three aspects of the evolution of $\beta$ in SS 316L, which have not yet been shown in prior work: $\beta$ is relatively insensitive to the applied stress during elastic deformation, $\beta$ can decrease with increasing plastic strain, and $\beta$ saturates with increasing plastic strain. Detailed discussions are provided to interpret these results in terms of dislocation structure evolution in stainless steel 316L under tensile loads.

2. Sample preparation and mechanical test

The hot-rolled stainless steel (SS) 316L block was machined into a dog bone specimen with electrical discharge machining (EDM). The as-received material was hot-rolled at 1100C and water-quenched for cooling. The yield stress of the as-received material was 425MPa. To reduce the yield stress to a similar level (~220MPa) reported in prior work[13,14], the material was heat-treated for solution annealing at 1100C for 1hour, followed by water quenching to homogenize grains and minimize residual stress from hot working. The specimen surface was prepared by mechanically grinding the oxide layer with sandpaper of up to 2000grit for a smooth surface to improve the signal-to-noise ratio (SNR) of ultrasonic measurements.

The specimen was incrementally loaded up to 6.0% total strain in stress control using the strain rate of 0.001 s$^{-1}$. The loading paths and the material behavior are provided in FIG 1: The sample was first loaded to 0.72% plastic strain, then unloaded to 0MPa, then loaded to 1.3% plastic strain, then unloaded, and this was repeated at 3.0% and 5.7% strain. Note that the unloading portion of the stress-strain curve is not plotted. The 0.2% yield stress of the material is 225MPa, and Young's modulus is 200GPa. The tensile test was interrupted at each data point in FIG 1 to hold the specimen at fixed stress levels for ultrasonic measurements. The markers in FIG 1 represents where ultrasonic measurements were made.

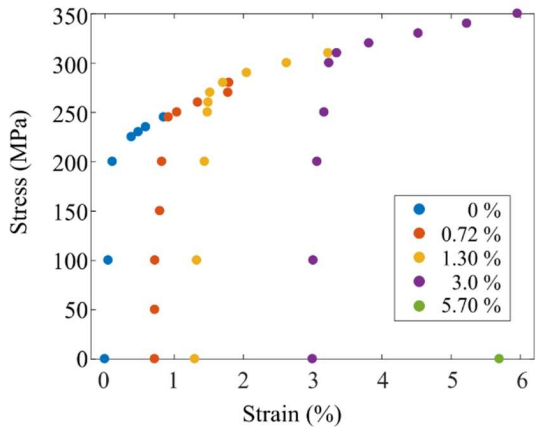

FIG 1. A stress-strain curve of the tested SS 316L specimen. The loading history consists of 4 segments, starting at a different plastic strain as noted in the legend. The markers represent where ultrasonic measurements were made. $\beta$ was measured three times when the applied stress is zero and only once when the applied stress is present

3. In situ nonlinear Rayleigh wave measurements

3.1. Overview and background of NLU

Nonlinear ultrasound is a technique to measure material nonlinearity. When an ultrasonic wave of a fundamental frequency ($f$) propagates through a material with defects, a portion of its energy transfers to the second harmonic ($2f$) due to the nonlinear stress-strain response. The acoustic nonlinearity parameter $\beta$ is a metric that quantifies the nonlinear ultrasonic response of defects. The relation between $\beta$ and the Rayleigh wave parameters is:

$$\beta^{rayleigh} = \frac{u_2}{xu_1^2} \frac{i8\sqrt{k_R^2 - k_L^2}}{k_L^2 k_R} \left(1 - \frac{2k_R^2}{2k_R^2 - k_S^2}\right), \tag{1}$$

where $u_1$ and $u_2$ are the vibration displacement amplitudes of the first ($f$) and the second ($2f$) harmonics respectively and $x$ is the propagation distance. The other variables are the wavenumber for a longitudinal wave ($k_L$), a shear wave ($k_S$), and a Rayleigh wave ($k_R$).

A relative acoustic nonlinearity parameter, $\beta^{rel}$, can be measured experimentally in two ways, as either

$$\beta_{dist}^{rel} \propto \frac{A_2}{xA_1^2} \text{ or } \beta_{amp}^{rel} \propto \frac{A_2}{A_1^2}, \tag{2}$$

where $\beta_{dist}^{rel}$ is used for distance sweep (i.e., $A_1$ is fixed) and $\beta_{amp}^{rel}$ is for amplitude sweep (i.e., $x$ is fixed). Here, $A_1$ and $A_2$ are the measured amplitudes corresponding to $f$ and $2f$ respectively. In this work, we used amplitude sweep for measurements and the detail is discussed in section 3.2. From here on, we use $\beta$ to refer to $\beta_{amp}^{rel}$.

The measured nonlinearity is a sum of the intrinsic material nonlinearity $\beta^{lat}$ and the nonlinearity of defects $\beta^{def}$, i.e., $\beta = \beta^{lat} + \beta^{def}$. The first term, $\beta^{lat}$, is defined as $\beta^{lat} = A_{111}/A_{11}$, where $A$ is the Huang's coefficients written in Voigt notation, and remains almost constant during plastic deformation[15]. The second term, $\beta^{def}$, which is the excess nonlinearity due to defects such as dislocations, continuously change during plastic deformation; thus, $\beta$ is used as a metric to evaluate the accumulation of dislocation-based damage.

The dislocation contributions to $\beta^{def}$ depend on the geometric configuration of dislocations, e.g., monopoles and dipoles. The models developed in prior work[2-4] have similar forms as shown in Eq. 3 and Eq. 4.

$$\beta_{mono}^{def}/\Lambda_{mono} b^2 = f_1(\nu) g_1(\Omega, R^3) \left(\frac{L}{b}\right)^4 \left(\frac{\sigma}{\mu}\right) \tag{3}$$

$$\beta_{di}^{def}/\Lambda_{di} b^2 = f_2(\nu) g_2(\Omega, R^2) \left(\frac{h}{b}\right)^3 + f_3(\nu) g_3(\Omega, R^3) \left(\frac{h}{b}\right)^4 \left(\frac{\sigma}{\mu}\right) \tag{4}$$

Here, subscript *mono* and *di* represent dislocation monopoles and dipoles respectively. The material constants such as Poisson's ratio ($\nu$), burgers vector ($b$), orientation factors ($\Omega$ and $R$), and shear modulus ($\mu$) are invariant during plastic deformation. $L$ and $h$ are monopole loop length and dipole height and $\Lambda$ is dislocation density. Finally, $\sigma$ is the internal stress acting on dislocations.

These equations readily show that both dislocation monopoles and dipoles have a linear dependence on the internal stress and that dislocation dipoles have an additional stress-independent term. In prior work, the internal stress term was accounted for in two different ways. Cantrell[2,16] and Apple[17] used the composite model of Mughrabi[18] and the endurance limit stress (the stress below which materials do not fail) to estimate the local stress state of dislocation substructures. In this approach, the internal stress evolution during plastic deformation is driven by the formation of veins and persistent slip bands (PSBs). The second approach uses the average residual stress from dislocation pile-ups to approximate the internal stress acting on dislocation substructures[19,20]. Compared to the first approach, the second approach is more useful for the planar slip metals, where veins and PSBs are hard to develop.

The internal stress calculated using two approaches is back stress generated by specific dislocation configurations. The applied stress in the macroscopic stress-strain curve is a sum of the frictional stress and the back stress[21,22]. The frictional stress represents the local resistance of the lattice to dislocation motions due to short-range interactions of dislocations. The back stress is induced by the long-range elastic interactions among dislocations due to grains and stress fields of dislocations. The evolution of the back stress depends on the number density of dislocations, which usually increases with plastic deformation. The advantage of using the back stress is that it relates to the macroscopic stress-strain curve. Thus, one can study the stress dependence of $\beta$ by changing the applied stress and measuring $\beta$ in situ during loading.

3.2. In situ measurement setup and method

We used a contact transducer of a 0.5" diameter as a transmitter and an air-coupled transducer (ACT) as a receiver. A sinusoidal burst of 40 cycles tuned to the center frequency of the transmitter (2.075MHz) was excited with Ritec RAM-5000 into the transmitter. The signal collected by the receiver was amplified into 30dB with a pre-amplifier. FIG 2 (a) shows the measurement configuration, and FIG 2 (b) shows the specimen after mounting ultrasonic fixtures and transducers. The representative time-domain signal is given in FIG 3 (a). The steady-state portion of the received signal was windowed with a Hanning window for FFT to extract the amplitude of the first (2.075MHz) and the second (4.15MHz) harmonics. The slope of $A_2/A_1^2$ is calculated as $\beta$ (FIG 3 (b)), and the same measurements were repeated three times to get the standard deviation of measurements. Specifically, $\beta$ was measured three times when the applied stress was zero and only once when the applied stress was non-zero.

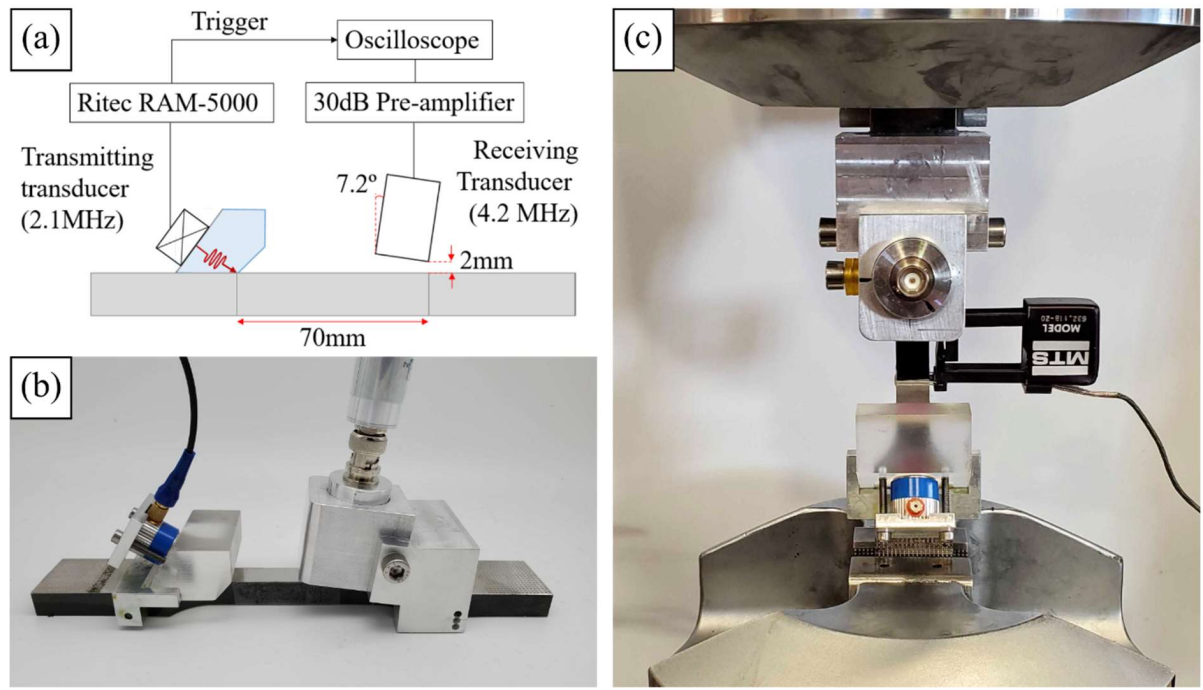

FIG 2 (a) *in situ* NLU Rayleigh wave measurement schematic. The Rayleigh wave generated by the contact transmitting transducer interacts with all the dislocation substructures in the entire gauge length (70mm) before it is collected by the air-coupled transducer. (b) The fixtures holding the wedge and the air-coupled transducer are glued to the specimen using silicone rubber. Note that the fixtures and the wedge are positioned outside the gauge length to prevent debonding during deformation. (c) The *in situ* measurement setup in the load frame. All the attachments remain mounted during *in situ* measurements except the wedge. The extensometer is mounted on the side of the specimen.

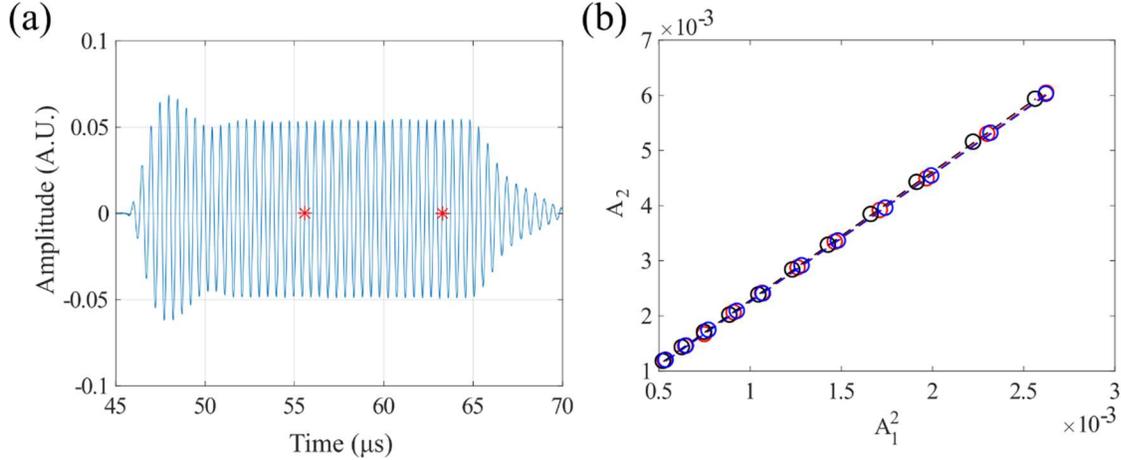

FIG 3 (a) A representative time domain signal. A sinusoidal burst of 2.075MHz was excited and the steady-state portion of the signal marked with red stars was used for post-processing to extract the amplitude of $A_1$ and $A_2$. (b) Markers with different colors represent three different amplitude sweeps. Transducers were remounted after each sweep to check the measurement error. The slope of each amplitude sweep is calculated using linear regression to obtain $\beta$.

As shown in FIG 3 (b), we used amplitude sweep in this work instead of distance sweep, which is more common[5,23], to calculate $\beta$. The advantage of distance sweep is that the system nonlinearity can be isolated since the output of the high-power amplifier is fixed[24]. Further, one can obtain more accurate results when using an ACT as a receiver by reducing a source of a measurement variation. However, distance sweep is sometimes not desirable if the geometry of the test specimen cannot be assumed as ideal, i.e., an infinite elastic half space. A few examples include geometries such as a curved surface[25,26] and a narrow dog-bone specimen[23]. For curved surfaces, a measurement variation is mainly due to the time-varying coupling of the wedge-specimen interface and difficulty with controlling the coupling layer[25]. In the case of a dog-bone geometry, the gauge width is usually comparable to the width of the effective line source (i.e., the wedge front). The boundary effect from the specimen is noted by Herrmann et al.[5], but currently, there is no available model to account for this phenomenon. In our preliminary measurements using distance sweep, we observed a periodic fluctuation of $A_1$ with a propagation distance, which significantly distorts the slope of the $A_2/A_1^2$ vs $x$. Thus, we adopted amplitude sweep to isolate the boundary effect by fixing the measurement position and considered the contribution from the system nonlinearity to $\beta$ as constant during plastic deformation.

3.3. Design considerations

We used two custom-made ultrasonic fixtures to constrain the position of the transducers (FIG.2 (a)). To determine the angle of the ACT fixture, we found the angle where $\beta$ is the maximum instead of using the theoretical value, i.e., $\sin^{-1}(c_{air}/c_{Rayleigh})$. The angle is 7.2degree and used instead of the theoretical value of 6.7degree. For the alignment of the transducers, a saddle-like shape was chosen while minimizing the clearance for pressure-fit to prevent slipping and rotation of fixtures. Similarly, the clearance between the wedge and the inner width of the wedge fixture was minimized for pressure fit.

For repeatable *in situ* Rayleigh wave measurements, two things must be considered: the location of ultrasonic fixtures

(or transducers) on the specimen and the near-field distance. The fixtures must be mounted on the outside of the gauge area to avoid decoupling of fixtures during loading. In our preliminary measurements, the silicone rubber easily debonded after 1-2% strain when the fixtures were on the gauge area. This is especially important to reduce the measurement variability from remounting the wedge. The gauge length of the tensile specimen must be determined based on Rayleigh distance to avoid the near-field effect (i.e., the fluctuation of the wave amplitude near the source). Rayleigh distance, $x_0$, for SS 316L and the frequency used (2.075MHz-4.15MHz) is 87.5mm and the near-field distance is 38mm ($0.435x_0$)[27]. The gauge length was designed to be 70mm, considering the size of ultrasonic fixtures and the extensometer.

3.4 Measurement procedures for *in situ* measurements of $\beta$

During *in situ* ultrasonic measurements, a specimen is mounted vertically in the load frame. We measured the time dependence of $\beta$ in the vertical and the horizontal orientations of the wedge to check the effect of time and orientation on the wedge-transducer coupling (FIG 4). Based on the result, the optimal procedure is to wait 40 minutes after the wedge-transducer assembly is mounted in the load frame.

The actual in situ $\beta$ measurements involves multiple steps: first, we glued the two fixtures to the specimen with silicone rubber, and pre-compressed them with C-clamps to cure for at least 12 hours. These fixtures were designed to hold the ACT and the wedge. This sample-fixture assembly was then mounted in the load frame as shown in FIG 2 (c) and waited 40 minutes. Next, we prepared wedge-specimen coupling by applying a gel couplant (Olympus D12) to the bottom of the wedge. Note that the gel couplant between the specimen and wedge was chosen to minimize couplant leakage due to gravity, while an oil couplant was used to couple the transducer-wedge interface to optimize ultrasonic energy transfer. The wedge was then *pressed fit* inside the fixture (left assembly in FIG 2 (b)) and compressed on the top manually to remove any empty space. The excessive couplant around the wedge front was carefully cleaned with a brush. While the transmitting transducer remained mounted on the wedge during the entire tension test, the wedge was remounted after each amplitude sweep. While this can cause variation in coupling condition and thus variation in the measurements, our repeated measurements show a measurement error of <1% (see Section 4). This indicates that the change in coupling condition is negligible.

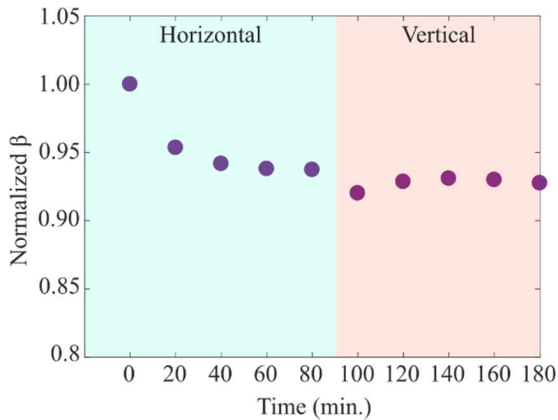

FIG 4 The dependence of $\beta$ on time and orientation. Transducers and the wedge were mounted only once for the first measurement (0 min.) and left untouched in the following measurements. Note that the specimen was initially horizontal (as shown in in FIG 2 (c)), then physically rotated 90degrees such that it was oriented vertically.

Lastly, we observed a 0.2-0.3% increase in total strain while the specimen is held at a given stress level, i.e., room temperature (RT) creep. However, RT creep seems to happen rapidly right after holding the load for ultrasonic measurements[28]. The change of dislocation substructures due to creep during ultrasonic measurements is not significant enough to influence the NLU measurements. FIG 5 shows that the amplitudes of $A_1$ and $A_2$ measured right after the initial amplitude sweep did not change much compared to those measured in the initial amplitude sweep. The difference in β between the initial amplitude sweep (black markers) and the modified amplitude sweep (the last two black markers replaced with red markers) is only 0.5%, which is significantly smaller than the changes due to different stress levels. The strain drop can be avoided by running strain-controlled tests, but then there would be stress-relaxation instead (i.e., a decrease in stress while strain is held)[28]. Thus, stress-controlled tests are still better to study the effect of the applied stress on $\beta$ at room temperature, which is the aim of this work.

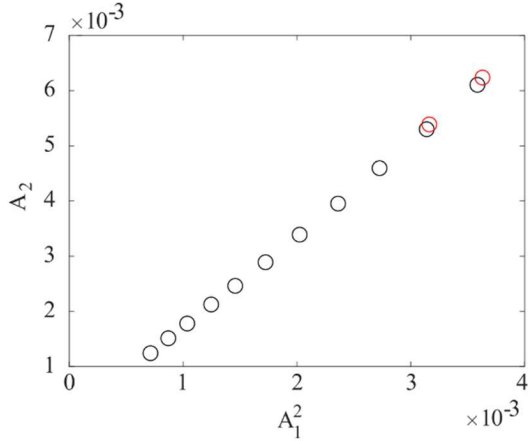

FIG 5 A representative *in situ* measurement. To check the effect of RT creep on the ultrasonic measurements, two additional data points (red) were collected after one amplitude sweep (black). The difference between $\beta$ was calculated from the initial amplitude sweep (11 black markers) and the modified amplitude sweep (9 black + 2 red markers) to quantify the effect of strain relaxation. The difference is only 0.5%, which is smaller than the measurement variation (<1%).

4. Results

FIG 6 shows the dependence of $\beta$ on plastic strain in the unloaded configuration (0MPa); note these measurements were conducted while the specimen was mounted in the load cell. Measurements show that in the undeformed state, $\beta$ is about 2.45, and it decreases rapidly to 1.77 at 1.30% plastic strain, followed by a saturation of $\beta$ upon further plastic deformation to 5.7% plastic strain. Three amplitude sweeps were repeated at each value of plastic strain at 0MPa to check measurement repeatability. As indicated by small error bars on FIG 6 (a), the standard deviation of our measurements is <1%. In FIG 6 (b), additional data points are plotted at each stress-strain condition as shown by the markers in FIG.1. The data points slightly deviate from the trend shown in FIG 6 (a) below 1.8% total strain (blue and dark orange circles), and do not alter the trend above 1.8% total strain (yellow and purple circles). The vertical deviation of data points from the red dotted line represents the effect of the applied stress. Below 1.8% total strain, the vertical difference is more significant than that above 1.8% total strain, which implies $\beta$ is more sensitive to the applied stress at low plastic strain. The trend of decreasing $\beta$ is particularly noteworthy, as all prior studies report that $\beta$ increases with plastic strain[2,5,20,23,29,30], which is in contrast to our results, both in situ during loading and at 0MPa load.

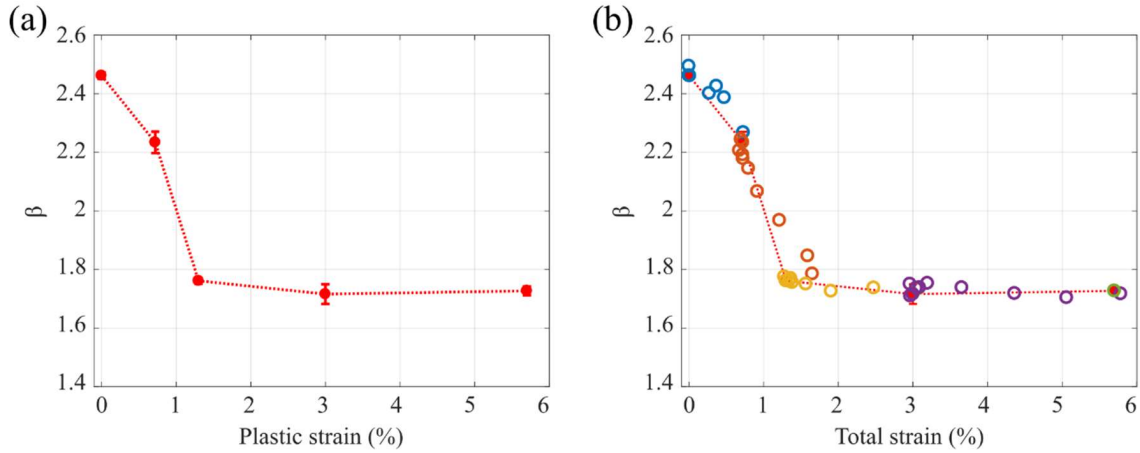

FIG 6 (a) Dependence of $\beta$ on plastic strain at 0MPa. Three repeated measurements of $\beta$ were made at 0MPa (see FIG 1) to check the standard deviation. The average of the standard deviations is <1%, which confirms the measurements are sufficiently repeatable. (b) The dependence of $\beta$ on total strain at all stress levels tested plotted in addition those at 0MPa from FIG 6(a). $\beta$ saturates after 1.8% total strain (or 125% yield stress). Each color represents the loading path shown in FIG 1.

To study the effect of the applied stress during elastic deformation, measured $\beta$ was plotted against the applied stress normalized by yield stress (FIG.7). For small plastic strain (0% and 0.72%), $\beta$ is more sensitive to the applied stress in the elastic regime; $\beta$ increases then decreases at 0% plastic strain, and $\beta$ decreases with the applied stress at 0.72% plastic stain. For larger plastic strain (1.3% and 3%), $\beta$ is relatively insensitive to the applied stress below the yield stress regardless of the amount of pre-deformation. The dependence of $\beta$ on the applied stress during elastic deformation can be quantified in terms of the slope of $\beta$ as a function of applied stress. The slope of $\beta$ is -0.0479 (blue) and -0.0853 (dark orange) for 0% and 0.72% plastic strains, and -0.0058 (yellow) and 0.0138 (purple) for 1.3% and 3%, respectively. Compared to smaller plastic strain, the slopes of $\beta$ for 1.3% and 3.0% plastic strains are smaller and closer to zero, implying that $\beta$ is more sensitive to the applied stress in early plastic deformation. These changes are not large, but they are not negligible. Further, $\beta$ clearly depends on applied stress in the plastic regime, and this dependence is independent of pre-deformation. However, the dependence of $\beta$ on applied stress beyond yield is also coupled to the dependence of $\beta$ on plastic strain.

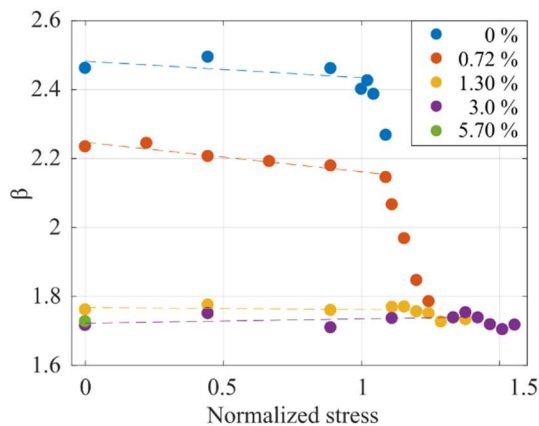

FIG 7 Dependence of $\beta$ on applied stress at different initial levels of plastic strain. Each color represents a different loading path as shown in FIG 1, and corresponds to the accumulated plastic strain before loading (or reloading). The applied stress is normalized by the yield stress (225MPa). $\beta$ is relatively insensitive to the applied stress in the elastic regime, rapidly decreases in the early plastic regime, and saturates around 1.8% plastic strain. The dashed lines

represent the least squares regression of β measured during elastic deformation to the linear function

5. Discussion

The results in FIG 6 and FIG 7 highlight three aspects of the evolution of β in stainless steel 316L: β is relatively insensitive to the applied stress during elastic deformation compared to plastic deformation, β decreases with plastic deformation, and β saturates at a particular plastic strain. To interpret these results, this section discusses theoretical models and prior work.

5.1 The effect of elastic stress on the model parameters and β

To explain the effect of applied stress on β, we first need to understand the theoretical models. Based on Cash's models, we note the following. First, β could increase then decrease depending on how $L$ and $h$ change with the progression of plastic deformation. In Eq.(3) and (4), dislocation density $\Lambda$ and back stress σ increase with plastic deformation while dislocation characteristic lengths $L$ and $h$ decrease. These model parameters are interdependent and how they scale during deformation depends on the initial microstructure and materials. Second, the back stress is the only parameter that changes β during elastic deformation. The changes in dislocation parameters ($\Lambda$, $L$, and $h$) during elastic deformation is negligible; thus, an increase in applied stress would only increase the back stress, which will change β in the same manner as in the previous plastic deformation. In other words, the trend of β during the reloading (elastic deformation) is determined by the changes in β during the previous plastic deformation. For example, in FIG 7, β decreases during early plastic deformation (blue markers, normalized stress>1). When the specimen was reloaded from 0.72% plastic strain (dark orange markers, normalized stress<1.1), β decreases in the elastic regime. Similarly, if β saturates during plastic deformation as shown for 1.3% or 3.0% plastic strain in FIG 6 (b) and FIG 7 (yellow and purple markers), β would not change much during the subsequent reloading. The changes to the slope of β in the elastic regimes shown in FIG 7 support this argument. When β decreases with plastic deformation (0% and 0.72% plastic strain) the magnitude of the slopes of β during the subsequent elastic deformation are an order of magnitude larger than when β remains constant with plastic deformation (1.3% and 3% plastic strain), where the slopes are closer to zero in the elastic range.

To understand the growing insensitivity of β to the back stress, we consider three different sets of model parameters to represent the progression of plastic deformation. The model parameters are assumed based on the range of parameters reported in prior work[21,31–33] as follows: $\Lambda=10^{13}$, $L=100$nm, $h=25$nm, and σ=120MPa, $\Lambda=10^{14}$, $L=77.5$nm, $h=20.2$nm, and σ=160MPa, and $\Lambda=5\times10^{14}$, $L=46$nm, $h=12$nm, and σ=200MPa. The values chosen reflect the interdependence of dislocation parameters: back stress and dislocation density increase with plastic deformation, and $L$ and $h$ decrease with increasing dislocation density due to more frequent dislocations interactions. The other parameters used for calculation are: $\Omega=R=0.33$, $v=0.31$, $b=0.295$nm and $\mu=76$GPa. In FIG 8 (a), the contributions from edge monopoles (black) and edge dipoles (red and blue) are plotted as a function of back stress using Cash's models[3,4] for three combinations of dislocation density and characteristic length defined above. The green horizontal line represents the lattice anharmonicity (i.e., inherent material nonlinearity), which is assumed based on the theoretical values for FCC metals[34]. FIG 8 (a) confirms that the contributions from dislocations can initially increase then decrease depending on the scaling of dislocation parameters as we argued at the beginning of this section. For example, we assumed that $h$ decreases from 25nm to 20.2nm, and then to 12nm as both dislocation density and back stress monotonically increase. The corresponding $\beta_{est}$ initially increases from 3.26 to 18.57 and then decreases to 14.47 (see asterisks on blue lines).

The total nonlinearity is as a sum of all contributions, i.e., the lattice anharominicity (green), monopoles (black), and dipoles (red and blue). Alternatively, β can be rewritten in terms of the stress-dependent and stress-independent terms, i.e., $\beta_{est}^{tot} = \beta_{est}^{lat} + \beta_{est}^{ind} + \beta_{est}^{dep}$. Here, $\beta_{est}^{lat}$ is the lattice anharmonicity (green), which is constant throughout plastic deformation, $\beta_{est}^{ind}$ is a stress-independent term (red), and $\beta_{est}^{dep}$ is a stress-dependent term (black and blue). To evaluate the change of the total nonlinearity $\beta_{est}^{tot}$, the data points corresponding to 120MPa, 160MPa, and 200MPa (see asterisks in FIG 9 (a)) were summed up, respectively, and plotted as a black dashed line in FIG 8 (b). $\beta_{est}^{tot}$ initially decreases and then saturates, which is qualitatively similar to the experimental results in FIG 6 (a). The red dashed line is $\beta_{est}^{dep}$, which describes the sensitivity of $\beta$ to the back stress. A negative $\beta_{est}^{dep}$ indicates that $\beta_{est}^{tot}$ would decrease as the back stress increases as other terms are constant. If $\beta_{est}^{dep}$ is close to zero, $\beta_{est}^{tot}$ would not change much with the back stress. Thus, the change of $\beta_{est}^{dep}$ from a negative value toward zero implies that $\beta_{est}^{tot}$ can become stress-insensitive as plastic deformation continues. This provides a possible explanation on why the slopes of the measured $\beta$ (dashed lines in FIG 7) become flat with plastic deformation. The significance of the similarity is that the stress-sensitivity of $\beta$ depends on the level of plastic strain and that the relative contributions from edge monopoles and dipoles determine the degree of sensitivity.

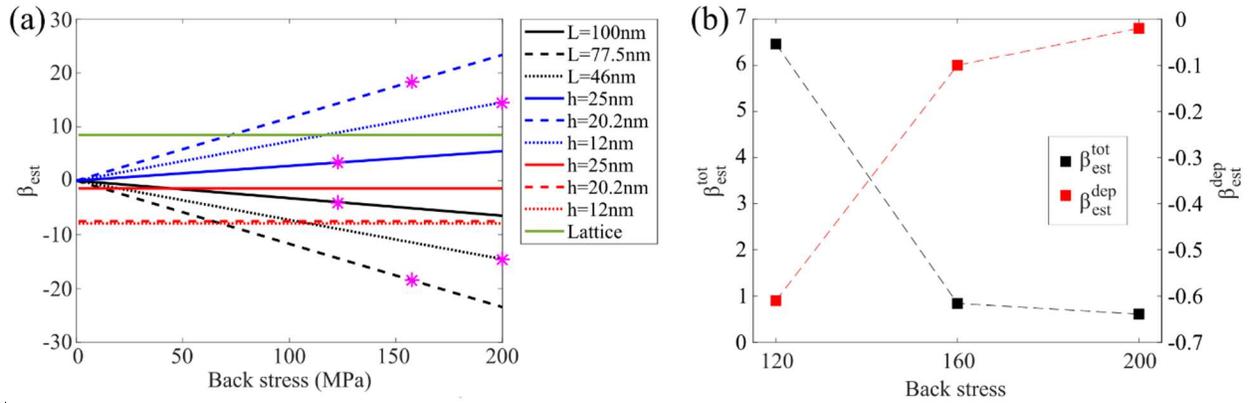

FIG 8 (a) Stress-dependence of β for three combinations of assume dislocation density: $\Lambda=10^{13}$ (solid lines), $10^{14}$ (dashed lines) and $5\times10^{14}$ (dotted lines). The contributions to β shown are: the lattice anharmonicity (green), edge monopoles (black), stress-dependent edge dipoles (blue), and stress-independent edge dipoles (red). (b) Dependence of $\beta_{est}^{tot}$ on back stress, calculated by summing all contributions in (a) at each stress level, compared to dependence of the stress dependent term $\beta_{est}^{dep}$ on back stress, calculated by summing the contributions from the starred data points in (a).

It should be highlighted that the dislocation parameters used for model predictions in FIG 8 are not measured experimentally. However, the model predictions show that it is *possible* for dislocations to change in such a way over applied stress to cause the dependence of β on applied stress as measured in FIG 7. These model predictions also show that the stress-sensitivity of β depends on competing effects among multiple dislocation parameters. A fully quantitative comparison requires systematic characterization of the evolution of dislocation structures and parameters, some of which could be characterized using transmission electron microscopy (TEM) and scanning electron microscopy (SEM). Also, probing more stress levels in the elastic regime with a low plastic strain level will help confirm how the change of β during elastic deformation depends on plastic deformation. We expect that β also depends on applied stress in the plastic region, but it is confounded with changes in dislocation parameters, which makes it hard to separate the effects.

## 5.2 Decreasing β and potential existence of negative β

The results in FIG 6 contrast those reported in prior work: β clearly *decreases* with increasing plastic strain. Based on the discussions in section 5.1, dislocation density, back stress, and dislocation characteristic lengths (h and L) drive the evolution of β. Changes to both dislocation density and back stress during tensile deformation always cause β to increase: dislocation density of well-annealed materials increases during tensile deformation to accommodate the imposed plastic strain, and the average back stress is proportional to the square root of dislocation density[20,31,35]. On the other hand, changes to dislocation characteristic lengths during tensile deformation will cause a decrease in *β*. Indeed, *L* and *h* decrease with plastic deformation; *L* decreases due to the generation of more pinning points by the interaction of dislocations in the primary gliding plane with immobile dislocations such as jogs or Lomer lock[36], and *h* tends to decrease as dislocation structures reorganize into dislocation-rich walls[33]. Thus, changes to *L* and *h* should attribute to the decrease of *β*.

Further, recent dislocation dynamics simulations show that monopoles and dislocation pileups can result in a negative *β*[3,11]. According to Cash's model[3], edge monopoles generate negative *β* if Poisson's ratio *v* is larger than 0.28, which could be the case for SS316L, and edge dipoles generate positive *β*. Thus, *β* could increase or decrease with the progression of plastic deformation depending on the relative magnitudes of *L* and *h*, since these parameters dictate the strength of the monopole and dipole terms in Eq. 3-4, respectively. If *L* is large enough or *h* is small enough, meaning the monopole term dominates, *β* could decrease with damage accumulation as shown by model predictions in FIG 8 (b).

According to the dislocation pile-up model, dislocation pile-ups always generate a negative *β* because the higher order elastic constants of dislocation pile-ups are negative[11]. In stainless steel 316L, dislocations preferably exist in the form of pile-ups[21,37]. This is because SS316L is a planar slip metal with a low stacking fault energy (SFE), where grain boundaries block the gliding motion of dislocations. An edge monopole can exist as a monopole while piled-up. Thus, the contribution from monopoles is not distinguishable from that of pile-ups. Experimentally measured *β* is the sum of the second harmonic amplitudes generated by monopoles, dipoles, and their structures. Thus, *β* could decrease depending on the contribution from dislocation pile-ups.

We should note that in theoretical models other than Cash et al., monopoles generate positive *β*[1,31,38]. However even in this case, *β* could decrease depending on how *L* changes during deformation relative to dislocation density, since *β* scales with $L^4$ but is only proportional to dislocation density. Regardless of the validity of existing theoretical models, the increasing/decreasing trend of *β* depends on the magnitude and initial value of *L*, which depends on initial microstructure.

5.3 Contrast to other work relating *β* to plastic strain

The measured decrease in *β* with plastic strain in FIG 6 is in strong contrast to every other work that measured *β* over increasing plastic strain. *β* was shown to increase with plastic strain in aluminum alloy[39], SS 304[35], Ni alloy[20], Ti alloy[30], and carbon steel[26]. It is typically assumed that *β* increases during tensile deformation, and our results provide important insight that this is not always the case. We hypothesize that these contrasting results are due to the differences in initial microstructures, such as precipitates, secondary phase, and grain size. These features contribute to the characteristic lengths of edge monopoles, dipoles, and pile-ups, all of which also have competing effects on *β*.

For example, precipitates serve as pinning sites, which decrease *L* and thus decrease *β*. Similarly, grain size affects *L* by producing more pinning sites (i.e., grain boundaries) and by enhancing dislocation interactions near grain boundaries[37,40]. In the vicinity of grain boundaries, multiple slips occur to accommodate the intergranular strain incompatibility. Dislocation interactions produce immobile dislocations, which are additional source of pinning sites. Grain size could affect the negative contribution from edge monopoles by restricting or facilitating the shortening of edge monopoles. Thus, the measured *β* may increase or decrease depending on the magnitude of negative *β* with respect to that of positive *β*.

To test our hypothesis that the dependence of *β* on plastic strain depends on the initial microstructure, we compare in situ NLU measurements on three specimens subjected to different heat treatments. The three specimens were machined from the same batch of material, and each had an "as-received" heat treatment of hot-rolled at 1100C and water quenched. FIG 9 compares sample in the "as-received" condition to two samples with additional heat treatments: (1) a specimen with an additional solution annealing at 1100C for 1hour followed by *water quenching*, and (2) a specimen with an additional solution annealing at 1100C for 1hour but *oven cooled*. These heat treatments were selected to induce different grain sizes and phases. Specifically, the average grain size in the as-received specimen is expected to be smaller than that in the specimens subjected to additional solution annealing. The specimen subjected to oven cooling is expected to have σ-phase, which will not be present when specimens are rapidly cooled. The results in FIG 9 support our hypothesis that the evolution of *β* with tensile load is strongly dependent on the initial microstructure. Another thing to note is that *β* saturates at *the same value* in the as-received specimen and the solution-annealed specimen. This result is particularly interesting, as it suggests a saturation value for *β* with plastic strain that is relatively independent of grain size, but certainly more measurements would be needed to confirm this effect.

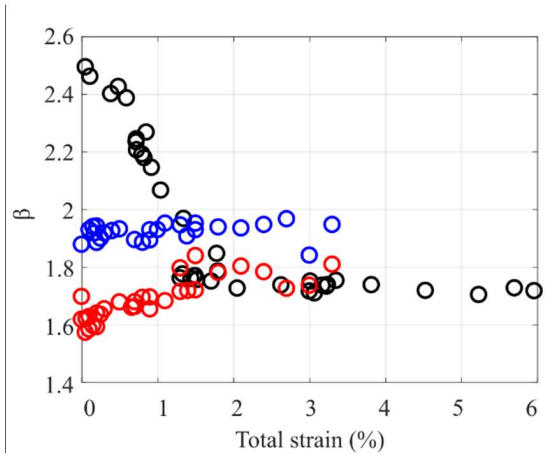

FIG 9 The comparison of the dependence of *β* on total strain measured in situ, for different heat treatments: an as-received sample, a sample with additional solution annealing and water quenching, and a sample with additional solution annealing but oven cooled.

However, why the specific material and heat treatment probed in FIG 6 exhibit decreasing *β* remains a question. Compared to prior work reporting increasing *β* on Ni alloy[20] (planar-slip), aluminum alloy[39] (wavy-slip), and SS 304[35] (similar composition), SS316L is a one-phase material (austenite) and precipitates are not expected from solution annealing and water-quenching. These differences in initial microstructures may lead to decreasing *β*. Unfortunately, the link between the initial microstructure and the evolution of β during plastic deformation has received very little attention, and many studies do not provide details on heat treatment or initial microstructure. Further studies on the relationship between initial microstructure (e.g., grain size) and the evolution of *β* and characterizing dislocation parameters using TEM could help explain the mechanism of decreasing *β*.

5.4 The saturation of *β*

The final important result in FIG 6 is the saturation of *β* at 1.8% strain. The rapid transition from the linear decrease (<1.8% strain) to the saturation (>1.8% strain) implies a significant change in dislocation substructure around this value of plastic strain. Feaugas reported that decreasing fractions of typical planar dislocations such as stacking faults and pile-ups coincides with the onset of multiple slips at 1.5% plastic strain in SS316L. Those planar dislocations transform to wavy-type dislocation structures such as edge dipoles, multipoles and tangles in SS 316L because of enhanced cross slip activity of screw dislocations due to multiple slip[21]. Among these wavy-type dislocations, edge dipoles and multipoles start to develop from the initial plastic deformation. At the time of the transition from planar to wavy slip, edge dipoles are observed in most grains and typical wavy-type dislocations such as tangles and walls are also formed. Considering edge dipoles and pile-ups (and edge monopoles) have competing effects on *β*, the saturation of *β* is a quantitative representation of the balance between these two structures. The constant *β* at higher plastic strains can be associated with the evolution of tangles and walls into cells. The formation of cell structures usually involves the saturation of dislocation parameters such as dislocation density and the fraction of screw and edge dislocations, all of which are due to the balance between dislocation generation and annihilation[18]. While we did not characterize dislocation structures in our specimen, it seems the dramatic transition in the trend of *β* from a monotonic decrease to saturation is associated with the transition from planar to wavy slip.

6. Conclusions

In this work, we developed an *in situ* Rayleigh wave measurement setup to understand the effect of the applied stress on the evolution of $\beta$ during plastic deformation. Using systematic procedures and custom-made fixtures, we were able to achieve repeatable in situ measurements, with maximum error of 1%. For *in situ* NLU Rayleigh wave measurements, a SS316L dog-bone specimen was incrementally loaded up to 6% total strain. The test was interrupted at given stress levels while loaded for ultrasonic measurements. The result shows that $\beta$ is relatively insensitive to the applied stress during elastic deformation and decreases with plastic strain. The apparent stress-insensitivity of $\beta$ was attributed to the competing effects between different dislocation configurations. Our results show that $\beta$ decreases with plastic deformation, which was not reported previously. We attribute this trend to the competing effects of dislocation parameters. If the initial characteristic length of dislocation structures, $L$, is sufficiently large from the heat treatment or manufacturing process, which is likely the case for this study, the negative contribution of $\beta$ is stronger and decreases $\beta$. Alternatively, $\beta$ could also decrease depending on the scaling of $L$ during plastic deformation without relying on the negative $\beta$. Follow-up measurements on samples with different heat treatments show that the trend of $\beta$ over plastic strain strongly depends on the initial microstructure. Lastly, $\beta$ saturated after 1.8% strain, which we posit is associated with the transition from planar slip to wavy slip. This work highlights three new aspects of the evolution of $\beta$ during tensile deformation: (1) the effect of the applied stress, (2) decreasing $\beta$ with increasing plastic strain accumulation, and (3) the saturation of $\beta$ with plastic strain. Further, the *in situ* NLU measurement setup and the procedures developed in this work will enable a more comprehensive understanding of how $\beta$ evolves in different materials over different damage mechanisms.

Acknowledgements

This material is based upon work supported by the National Science Foundation under Grant No. CMMI-20-15599. Experimental measurements were carried out in part in the Advanced Materials Testing and Evaluation Laboratory, University of Illinois at Urbana-Champaign.